\documentstyle[11pt,newpasp,twoside]{article}
\def\edcomment#1{\iffalse\marginpar{\raggedright\sl#1\/}\else\relax\fi}
\reversemarginpar

\begin{document}
\title{History of Astrophysics in Antarctica - A Brief Overview}
\vspace{-22pt}
\author{Balthasar T. Indermuehle$^\dag$, Michael G. Burton$^\ddag$, Sarah T. 
Maddison$^\dag$}
\affil{$^\dag$Swinburne University, Centre for Astrophysics and Supercomputing,
Hawthorn, VIC 3122, Australia}
\affil{$^\ddag$University of New South Wales, School of Physics,
Sydney, NSW 2052, Australia
}
\vspace{8pt}
\par
\noindent
On examining the historical development of astrophysical science at
the bottom of the world from the early 20th century until today we
find three temporally overlapping eras of which each has a rather
distinct beginning. These are the eras of Astrogeology, High Energy
Astrophysics and Photon Astronomy.\footnote{A full version of this
paper will be submitted to {\it PASA}. The poster is on the SPS2
website.}

\vspace{-11pt}
\section{Astrogeological Era (from 1912)}
\vspace{-6pt}
In 1912, Mawson discovered the Adelie Land Meteorite. Russian
geologists found several meteorites of unrelated morphological
types in the Lazarev region in 1961, giving rise to the ablation zone
theory. This was followed in 1969 by the first formal meteorite search 
programme.  Thousands of meteorites have since been found in Antarctica, 
mostly from glacial ablation zones.
%

\vspace{-11pt}
\section{High Energy Era (from mid 1950's)}
\vspace{-6pt}
In 1955, the first Antarctic astrophysical
project was initiated with a cosmic ray detector installed at McMurdo station. A detector was installed 
at the South Pole in 1964 and others were commissioned at several national
research stations. Today, cosmic rays with energies above 50 TeV are
recorded by {\it SPASE 2} which is situated on top of {\it
AMANDA}. {\it AMANDA II} is scheduled to start operation in 2003 and
by 2006, {\it IceCube}, a 1 km$^3$ detector, will open up the PeV
energy region where the Universe is opaque to high energy
$\gamma$-rays from beyond our galaxy.

\vspace{-11pt}
\section{Photon Astronomy Era (from 1970's)}
\vspace{-6pt}
Optical astronomy was first employed at the South Pole in 1964 for site
testing. In 1979, a continuous 120 hours of solar observations
allowed hundreds of solar eigenmodes to be discovered. Infrared and
CMB measurements require the best atmospheric windows, and site
testing since 1994 has demonstrated this is provided from the
Antarctic plateau. In 1988, the first CMB measurement from Terra Nova
Bay produced a map at an angular scale of 1.3$^{\circ}$. In 1992, the
CMB was imaged with better resolution by {\it Python}, a 0.75m
telescope. It was replaced in 1997 by {\it Viper}, a 2.1m off-axis
telescope, to be complemented in 1998 by DASI. The first flight of the
{\it Boomerang} balloon in 1998 provided much improved angular
resolution data for a CMB map published in late 2002. AST/RO, a 1.7m
sub-mm telescope, has been mapping the Galactic plane in C\small I \normalsize
and CO since 1995. Between 1994 and 1999, {\it SPIREX} imaged the 
2--5$\mu$m continuum and PAHs with the {\it Grim} and {\it Abu}
cameras.

\end{document}